\documentclass[sigconf,10pt]{acmart}

\usepackage{subcaption}
\usepackage{booktabs}
\usepackage{multirow}

\AtBeginDocument{%
  }

\acmYear{2026}\copyrightyear{2026}
\setcopyright{cc}
\setcctype[4.0]{by}
\acmConference[MobiCom '26]{The 32nd Annual International Conference on Mobile Computing and Networking}{October 26--30, 2026}{Austin, TX, USA}
\acmBooktitle{The 32nd Annual International Conference on Mobile Computing and Networking (MobiCom '26), October 26--30, 2026, Austin, TX, USA}
\acmDOI{10.1145/3795866.3844758}
\acmISBN{979-8-4007-2505-0/26/10}

\begin{document}

\title{WiP: Meter-Level Wi-Fi RTT Localization on a Production Enterprise WLAN}

\renewcommand{\thefootnote}{\fnsymbol{footnote}}

\author{Enguang Fan}
\authornotemark[1]
\affiliation{%
    \institution{University of Illinois Urbana-Champaign}
    \city{}
    \state{}
    \country{}}
\email{enguang2@illinois.edu}

\author{Binh Minh Tran}
\authornote{Both authors contributed equally to this research.}
\affiliation{%
    \institution{University of Illinois Urbana-Champaign}
    \city{}
    \state{}
    \country{}}
\email{bmtran2@illinois.edu}

\renewcommand{\thefootnote}{\arabic{footnote}}

\author{Klara Nahrstedt}
\affiliation{%
    \institution{University of Illinois Urbana-Champaign}
    \city{}
    \state{}
    \country{}}
\email{klara@illinois.edu}

\begin{abstract}

Wi-Fi Fine Time Measurement (FTM) promises indoor localization by reusing access points (APs) already deployed for connectivity, but prior evaluations mostly use APs purpose-deployed or calibrated for ranging, leaving it unclear whether a production enterprise WLAN can provide useful localization without localization-specific infrastructure. We evaluate Wi-Fi round-trip time (RTT) localization on IllinoisNet, a live campus WLAN whose APs were placed for coverage and capacity. Using five commodity Android phones at 10 static locations across a 50 m × 30 m office floor spanning LOS, NLOS, and multipath conditions, weighted nonlinear least squares achieves a median error of 1.21 m and a 90th-percentile error of 2.68 m—with no AP replacement, repositioning, or ranging calibration. Residuals vary in magnitude and sign across APs and propagation conditions, suggesting a single global correction is insufficient and motivating AP-aware software calibration rather than new localization-specific infrastructure.

\end{abstract}


\maketitle

\section{Introduction}

Wi-Fi Fine Time Measurement (FTM) is attractive for indoor positioning because it estimates AP--client distance from round-trip time (RTT) measurements while reusing infrastructure already deployed for network access~\cite{kosekszott2025indoorpositioningwifilocation}. In principle, this reuse could enable indoor positioning without installing a separate localization network.

In practice, however, the infrastructure-reuse premise remains insufficiently evaluated. Prior RTT localization studies commonly rely on temporary AP deployments or specialized routers configured or calibrated for ranging. Their APs can be selected and positioned for favorable ranging geometry, whereas a production enterprise WLAN is designed primarily for coverage and capacity. Such a deployment may therefore exhibit sparse or unfavorable AP geometry, together with non-line-of-sight (NLOS) and multipath propagation. Whether such an existing deployment can provide useful RTT localization therefore remains an open practical question.

We test this premise directly, evaluating Wi-Fi RTT on IllinoisNet, a live campus WLAN, using commodity Android phones, without AP replacement, repositioning, or ranging calibration. Two findings follow: meter-level indoor positioning is achievable without localization-specific infrastructure; and ranging residuals vary in magnitude and sign across APs and propagation conditions, so a single global bias correction is unlikely to suffice.

\vskip -4pt
\section{Testbed Setup}

\begin{figure}[t]
    \centering
    \begin{subfigure}{0.48\columnwidth}
        \centering
        \includegraphics[width=\linewidth]{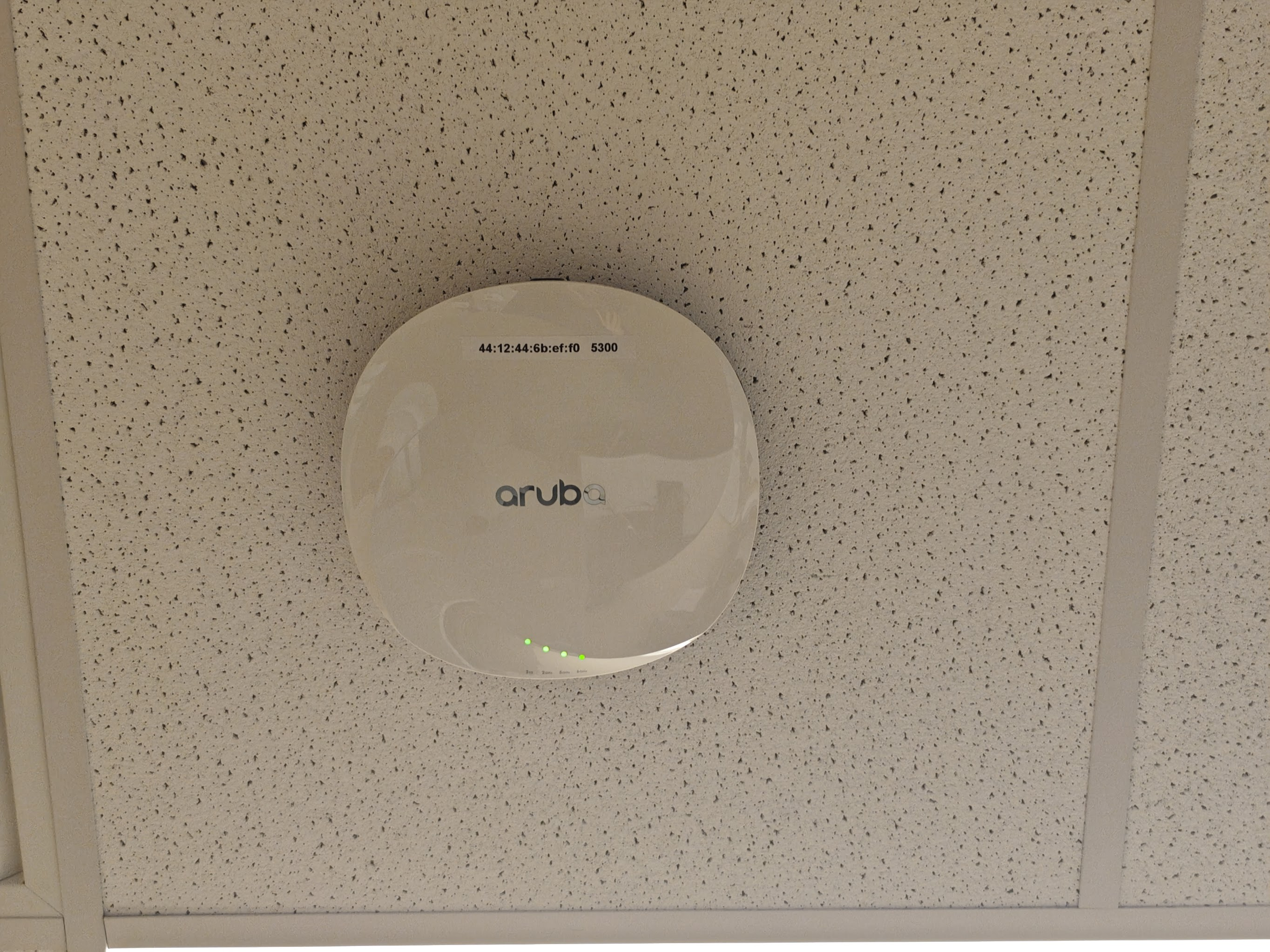}
        \caption{Aruba AP-635 access point serving the campus network.}
        \Description{Photograph of an Aruba AP-635 access point used in the IllinoisNet deployment.}
        \label{fig:ap-image}
    \end{subfigure}
    \hfill
    \begin{subfigure}{0.48\columnwidth}
        \centering
        \includegraphics[width=\linewidth]{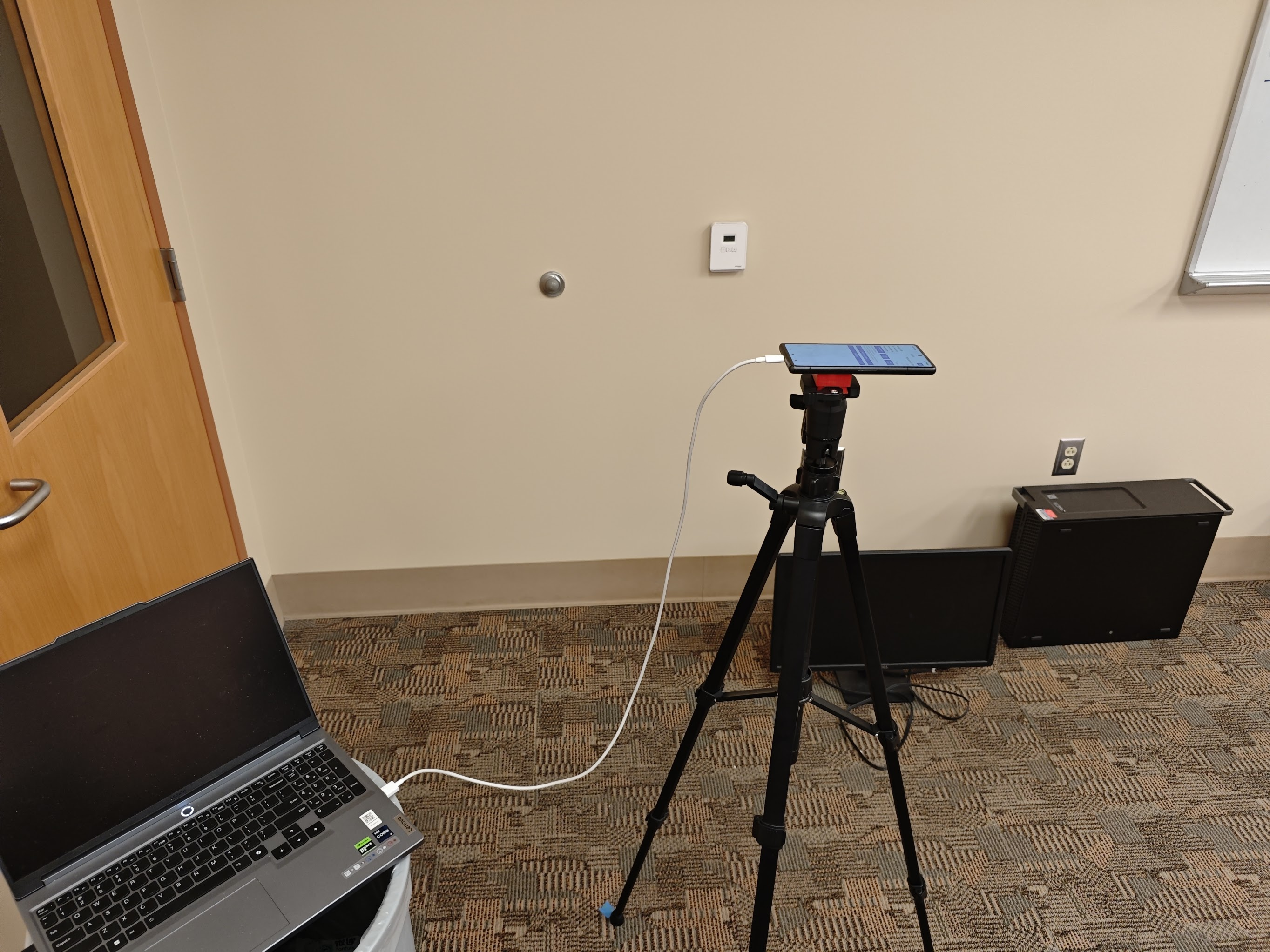}
        \caption{Phone mounted on a tripod for data collection.}
        \Description{Photograph of the phone mounted on a tripod for Wi-Fi RTT data collection.}
        \label{fig:phone_collector}
    \end{subfigure}
    \caption{RTT-capable AP and phone-based collection setup.}
    \label{fig:collection_system}
\end{figure}

\begin{figure}[t]
    \centering
    \includegraphics[width=0.55\columnwidth]{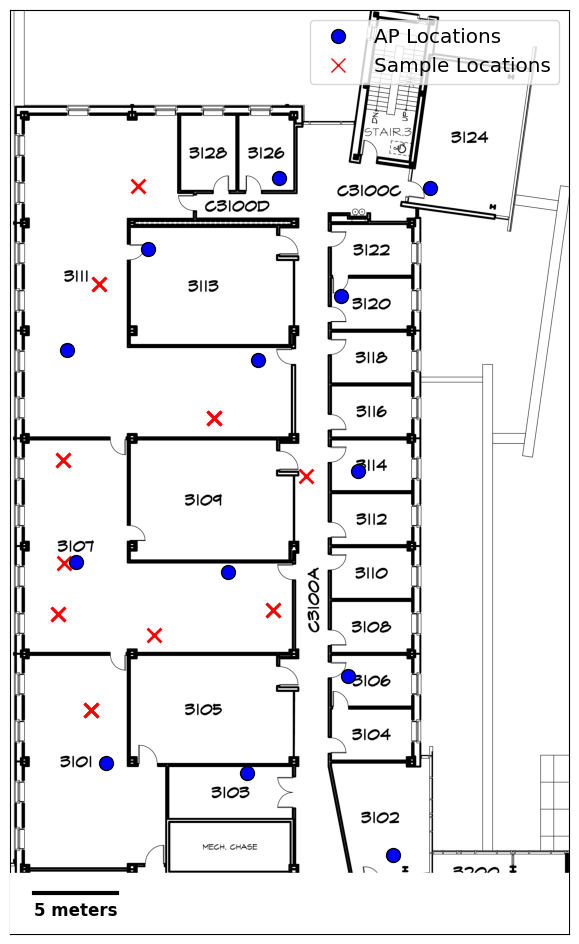}
    \caption{Testbed floor map with IllinoisNet APs and sampling locations.}
    \Description{Floor-map view of the left-wing testbed area with marked RTT-capable AP positions and static ranging sample locations.}
    \label{fig:env}
\end{figure}

\noindent\textbf{Environment and Propagation Conditions: }We conducted the evaluation in a university building where each floor spans approximately 105~m by 62~m. Our testbed covers the 50~m by 30~m left wing of one floor, as shown in Figure~\ref{fig:env}. The environment exposes RTT ranging to two qualitatively different sources of error. First, drywall partitions obstruct the direct path and introduce an NLOS-dependent ranging offset. Second, large reflective surfaces---including whiteboards, metal cabinets, lockers, and pipe covers---create multipath. Under strong multipath, a receiver may resolve a longer reflected path instead of the direct path, adding excess propagation delay to the measured RTT and increasing variability. These two propagation mechanisms motivate the ranging analysis in Section~\ref{sec:results}.

\begin{table}[t]
    \centering
    \footnotesize
    \begin{tabular}{@{}llrrr@{}}
    \toprule
    \textbf{Band} & \textbf{Channel} & \textbf{Width} & \textbf{Obs.} & \textbf{Share} \\
    \midrule
    \multirow{3}{*}{2.4~GHz}
       & 1  & 20~MHz & 3,446 & 13.2\% \\
       & 6  & 20~MHz & 3,441 & 13.2\% \\
       & 11 & 20~MHz & 3,496 & 13.4\% \\
    \addlinespace[2pt]
    \multirow{5}{*}{5~GHz}
       & 42  & 80~MHz & 3,878 & 14.9\% \\
       & 58  & 80~MHz & 3,127 & 12.0\% \\
       & 106 & 80~MHz & 3,338 & 12.8\% \\
       & 122 & 80~MHz & 2,778 & 10.7\% \\
       & 155 & 80~MHz & 2,550 & 9.8\% \\
    \midrule
    \multicolumn{3}{@{}l}{\textbf{Total}} & \textbf{26,054} & \textbf{100.0\%} \\
    \bottomrule
    \end{tabular}
    \caption{Operating-channel diversity observed in the IllinoisNet scan log. The 5~GHz primary channels are grouped by 80~MHz center channel; shares are computed across all scan observations.}
    \label{tab:IllinoisNet-channels}
\end{table}

\noindent\textbf{Production WLAN Infrastructure: }The building is served by IllinoisNet, the live campus WLAN, which uses Aruba AP-635 access points such as the one shown in Figure~\ref{fig:ap-image}. Because the deployment is designed for connectivity, AP placement prioritizes coverage and capacity rather than geometric diversity for localization. The resulting geometry is sparse and includes AP--client paths obstructed by walls and reflectors.

IllinoisNet also exhibits the channel diversity and dynamics of an enterprise WLAN. As is common in enterprise deployments, a centralized controller distributes APs across the 2.4~GHz and 5~GHz bands and may reassign their channels as interference conditions change. This differs from many controlled RTT evaluations, which keep APs on one fixed channel or a small set of known channels~\cite{kosekszott2025indoorpositioningwifilocation}. Our scan log contains the three standard 2.4~GHz channels and five 80~MHz groups in the 5~GHz band, as summarized in Table~\ref{tab:IllinoisNet-channels}. Our ranging system makes no fixed-channel assumption: it obtains each AP's current channel from ordinary scan metadata and constructs the FTM request using that current responder configuration. Controller-driven channel changes therefore require no manual reconfiguration.

\noindent\textbf{Accessing Unadvertised FTM Support: }Although the AP-635 firmware supports FTM and responds to ranging exchanges, IllinoisNet does not advertise the FTM-responder capability in its beacon frames. Consequently, standard Android discovery does not identify these APs as ranging responders. Our system addresses this discovery limitation by explicitly supplying the AP responder information from the scan result and issuing a two-sided FTM request. This procedure changes neither AP firmware nor network configuration; the client merely requests a capability already implemented by the deployed AP.

\noindent\textbf{Collection System: }We collected ranging data from commodity Android phones---Google Pixel 4a, 6a, 7, 7a, and Samsung Galaxy S21---using the setup shown in Figure~\ref{fig:phone_collector}. We selected 10 static sampling locations across the testbed and mounted each phone on a tripod at a height of 1.25 m to reduce measurement variability. At each location, each phone collected RTT measurements from all visible IllinoisNet APs for 30 seconds, yielding 50 measurements per location.

\section{Results}
\label{sec:results}

\begin{figure*}[!t]
    \centering
    \begin{subfigure}{0.49\textwidth}
        \centering
        \includegraphics[width=0.78\linewidth]{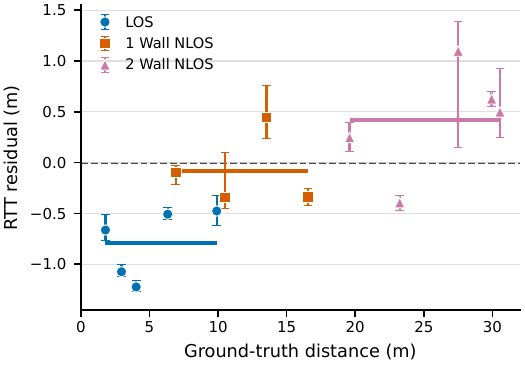}\\[4pt]
        \includegraphics[width=0.88\linewidth]{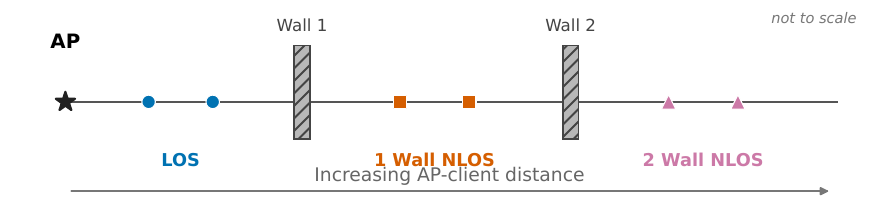}
        \caption{Drywall residual shift.}
        \label{fig:structured-errors-wall}
    \end{subfigure}
    \hfill
    \begin{subfigure}{0.49\textwidth}
        \centering
        \includegraphics[width=0.78\linewidth]{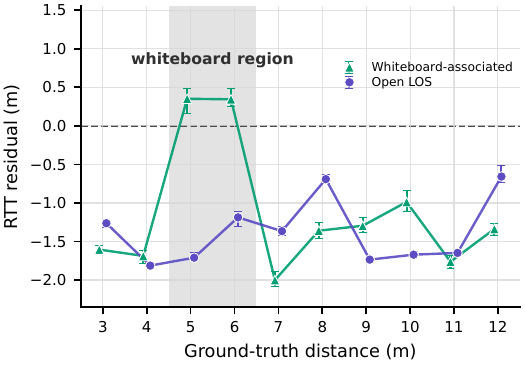}\\[4pt]
        \includegraphics[width=0.88\linewidth]{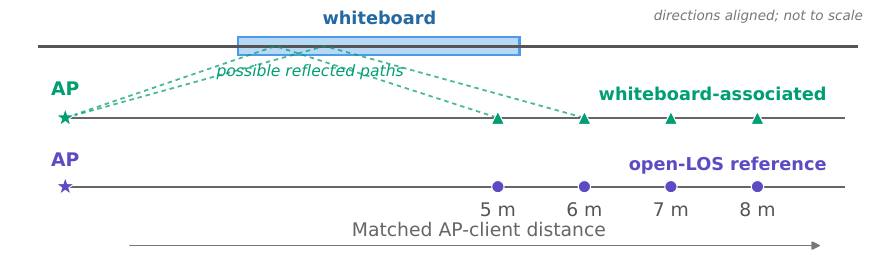}
        \caption{Whiteboard residual variability.}
        \label{fig:structured-errors-whiteboard}
    \end{subfigure}
    \caption{Structured RTT errors in a production WLAN. (a) Drywall-obstructed paths shift residuals upward. (b) A whiteboard-associated direction has \(2.05\times\) the median per-distance IQR of an open-LOS reference and a larger IQR at 10 of 12 distances. Schematics are not to scale.}
    \Description{Two-panel comparison of structured Wi-Fi RTT errors, with residual distributions and schematics for drywall obstruction and whiteboard-associated multipath.}
    \label{fig:structured-errors}
\end{figure*}

\begin{figure}[t]
    \centering
    \includegraphics[width=\columnwidth]{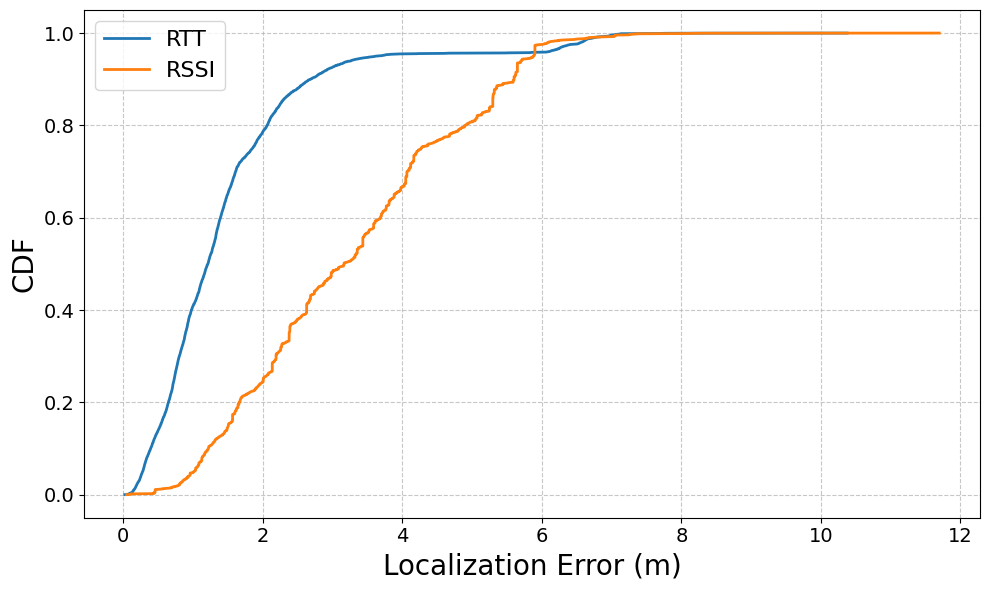}
    \caption{CDF of localization error for RTT and the RSSI multilateration baseline across all 10 test locations.}
    \Description{CDF plot summarizing localization error for RTT- and RSSI-based localization.}
    \label{fig:cdf}
\end{figure}

\noindent\textbf{Meter-Level Localization: }For each ranging epoch, we estimate the 2D client position using weighted nonlinear least squares:
\begin{equation}
\hat{\mathbf p}=\arg\min_{\mathbf p}\sum_{i\in\mathcal A}\frac{1}{|\rho_i|}\left(\|\mathbf p-\mathbf a_i\|_2-r_i\right)^2,
\label{eq:wnls}
\end{equation}
where \(\mathbf a_i\), \(r_i\), and \(\rho_i\) are the position, RTT-derived range, and RSSI in dBm for AP \(i\), respectively. Thus, a weak measurement such as \(-90\)~dBm receives less weight than a stronger one. As a baseline, we use multilateration with an RSSI log-distance path-loss model.

Figure~\ref{fig:cdf} answers our central question: the production WLAN provides meter-level static positioning despite AP placement that was not designed for localization. RTT achieves a median error of 1.21~m and a 90th-percentile error of 2.68~m, compared with 3.17~m and 5.60~m for the RSSI baseline, respectively.

\noindent\textbf{Structured Ranging Errors: }We define the ranging residual as \(e_i=r_i-d_i\), where \(d_i\) is the ground-truth AP distance. Figure~\ref{fig:structured-errors} separates two recurring effects. Drywall-induced NLOS shifts the residual upward as the direct path is obstructed, whereas a nearby whiteboard produces a wider distribution and more positive outliers through delayed reflected paths. The magnitude of these effects varies across APs, suggesting that a single global correction may be insufficient and motivating AP-aware calibration.

\section{Conclusion and Future Work}

Our evaluation shows that Wi-Fi RTT can provide practical indoor localization in a sparse, unoptimized deployment. The ranging analysis reveals AP-specific RTT offsets associated with distance and propagation conditions, motivating calibration methods that improve ranging accuracy without requiring localization-specific AP deployments. Our evaluation relies on manually surveyed AP coordinates; a future system could incorporate automatic AP-location estimation~\cite{EnguangPoster, sie2025crowdsourcingubiquitousindoorlocalization}.

\begin{acks}
This work was supported by the National Science Foundation under Grants NSF CCF 22-17144, NSF CNS 24-37204.
\end{acks}

\enlargethispage{7\baselineskip}
\begingroup
\scriptsize
\setlength{\bibsep}{0pt}
\bibliographystyle{ACM-Reference-Format}
\bibliography{bibliography}
\endgroup

\end{document}